\documentclass[preprint,showpacs,aps,floatfix]{revtex4-1}
\usepackage{graphicx}
\usepackage{graphics}
\usepackage{amssymb}
\usepackage{amsmath}
\usepackage{amsfonts}
\usepackage{tikz}
\usetikzlibrary{decorations.pathmorphing,decorations.markings}
\def\by#1#2{{\displaystyle {#1}\over \displaystyle {#2}}}
\def\d{{\rm d}}

\def\bea{\begin{eqnarray}}
\def\eea{\end{eqnarray}}
\def\mc{\mathcal}
\def\nn{\nonumber}

\def\al{\alpha}
\def\bt{\beta}
\def\g{\gamma} 
\def\dl{\delta}

\def\tta{\theta}

\def\lm{\lambda}

\def\s{\sigma}

\def\rh{\rho}
\def\om{\omega} 
\def\ph{\phi}

\begin{document}
\title{ Vector meson fragmentation using a model with broken
${{SU(3)}}$ at the Next-to-Leading Order}
\author{H. Saveetha$^1$, D. Indumathi$^2$, Subhadip Mitra$^3$}
\bigskip{}
\affiliation{
${}^1${\it Department of Theoretical Physics, University of Madras,
Chennai 600 025, India.} \\
${}^2${\it Institute of Mathematical Sciences, Chennai 600 113, India.}
\\
${}^3${Laboratoire de Physique Theorique d'Orsay, 
Univ. Paris-Sud 11, France.}}

\date{\today} 

\begin{abstract}

A detailed study of fragmentation of vector mesons at the next-to-leading
order (NLO) is given for $e^+ \, e^-$ scattering. A model with broken
$SU(3)$ symmetry uses three input fragmentation
functions $\al(x, Q^2)$, $\bt(x,Q^2)$ and $\g(x,Q^2)$ and a strangeness
suppression parameter $\lm$ to describe all the light quark
fragmentation functions for the entire vector meson octet. At a starting low energy scale of $Q_0^2
= 1.5$ (GeV)$^2$ for three light quarks $(u, d, s)$ along with initial
parameterization, the fragmentation functions are evolved through DGLAP
evolution equations at NLO and the cross-section is calculated. The
heavy quarks contribution are added in appropriate thresholds during
evolution. The results obtained are fitted at the momentum scale of
$\sqrt{s}= 91.2$GeV for LEP and SLD data. Good-quality fits are
obtained for $\rho$, $K^*$, $\omega$ and $\phi$ mesons, 
implying the consistency and efficiency of this model that explains
the fragmentation functions of vector mesons both at the leading and
the next to leading order in QCD.


\end{abstract}


\maketitle

\section{Introduction}

The fragmentation of partons (quarks,gluons) into the desirable hadrons
is called hadronization or fragmentation and are expressed
in terms of functions named fragmentation functions. Fragmentation
process can be understood only through phenomenological studies as
perturbative Quantum Chromodynamics (pQCD) itself cannot explain it in
a direct way. Various phenomenological studies (\cite{Bour}, \cite{Kret},
 \cite{Hirai}, \cite{Flor}, \cite{Strat}) are being done for many years
to understand the fragmentation process ( $e^+ \, e^-$, $e \, p$, $p \,
p$, lepton-nuclues,\ldots) at the leading order (LO) level and the next
to leading order(NLO) level for pions and kaons (pseudoscalar)as well
as baryons.

However, while considering vector mesons there is no such work has
been found so far due to the paucity of the data. This motivated
us to study the fragmentation of vector mesons. A model with broken
$SU(3)$ \cite{Indu},\cite{Misra} is used in order to explain these
fragmentation functions at the leading order level intially for $e^+ \,
e^-$ annihilation and  $p \, p$, collision processes \cite{Savee}. As a
consequence, it is aimed to extend the analysis for these vector mesons
at the next to leading order using the same model with broken $SU(3)$. The
significance of studying these vector mesons at the next-to-leading order
level, particularly for the $\phi$ meson, will be a guideline for sure
to understand the nucleon-nucleon scattering in the RHIC for Quark Gluon
Plasma (QGP) studies. This work presents the study of fragmentation
functions for $e^+ \, e^-$ at NLO.

Section $2$ of the paper explains the cross section details of $e^+\,e^-$
scattering at the Next to leading order level. Section $3$ describes
the outline of the broken $SU(3)$ model. Section $4$ gives the initial
parameterization.  Section $5$ using the model presents the expressions
of fragmentation functions of entire meson nonet. Section $6$ analyses
the results obtained by comparison with the data and discuss them
in detail.  Section $7$ concludes the work.

\section{Kinematics} 
\subsection{ Hadron production at the Next to
leading order level}

In this paper, the fragmentation of a parton into a colorless hadron
is being studied for $e^+\,e^-$ annihilation. The basic differential
cross-section of this process has the following form \cite{Fur}:
\bea
\frac{d\s^{h}_{e^+e^-}(x,y;Q^2)}{dxdy} = N_c\frac{4\pi\al^2_{\rm e.m.}}{3Q^2}
\left[ \frac{3}{2}f^h_{B,1}(x,Q^2) - 3y(1-y)f^h_{B,2}(x,Q^2) + \frac{3}{2}(1-2y)f^h_{B,A}(x,Q^2)\right] \nn \\
\label{eq:crossec}
\eea
 where 
\bea
Q^2 &=& (l_1+l_2)^2,\\
x   &=& \frac{2h\cdot (l_1+l_2)}{Q^2},\\
y   &=& \frac{h\cdot l_1}{h\cdot (l_1+l_2)}\\
    &=& \frac{1}{2}[1-\cos\tta_{\rm CM}],\\
B   &=&  \hbox{gauge boson} (g, Z  or  W). \nn
\eea
Here $x$ is the fraction of energy carried over by the hadron from its
parent quark, where $x \equiv E_{hadron}/E_{quark} =(2E_{h}/\sqrt{s})
\leq1$ ( or $x_p \equiv 2p_h/\sqrt{s}$) and Q=$\sqrt{s}$ is the energy
scale, where the analysis is done.Hence, after integrating over  y
($f^h_{B,A}$ term goes to zero), Eq.~[\ref{eq:crossec}] with, for
example, $Z$ boson as the propagator becomes

\bea
\frac{d\s^{h}_{e^+e^-}(x;Q^2)}{dx} &=& \left(\frac{N_c4\pi\al^2}{3Q^2}
\right)
\left[ \frac{3}{2}f^h_{Z,1}(x,Q^2) - \frac{1}{2}f^h_{Z,2}(x,Q^2)\right], \\
f^h_{Z,r}(x,Q^2)
         &=& \sum_P \mc C_r^P(x,Q^2)\otimes\lm^{P}_{Z} D^h_P(x,Q^2), \ r=1,2 \\
\lm_{Z}^{F} &=& {\lm_Z^{q_f}} =\lm_Z^{\bar q_f}
 = c_{q_f},\\
\lm_Z^G &=& \sum_{f} \lm^{F}_{Z}
\eea
where $D^h_p(x,Q^2)$ is the fragmentation function for a parton ($p$) that
fragments into one of the hadrons ($h$) which we are interested in. The
expressions for charge factors $c_{q_f}$ of each quark of flavor $f$
in terms of electromagnetic charge $e_f$, vector and axial vector
electroweak couplings are given in \cite{Savee}. The coefficient
functions, $\mc C_r^P(x,Q^2)$, are expressed in series of $\al_s(Q^2)$:

\bea
\mc C_r^F(x,Q^2) &=& \dl(1-x) + \frac{\al_s(Q^2)}{2\pi}\mc C_r^{F(1)}(x) + \ldots,\\
\mc C_r^G(x,Q^2) &=& \frac{\al_s(Q^2)}{2\pi}\mc C_r^{G(1)}(x) +\ldots~,
\eea
where $\mc C_r^{P(i)}$'s are now independent of $Q^2$. 
Hence at NLO,
\bea
f^h_{B,r}(x,Q^2) &=& \left(\dl(1-x)+\frac{\al_s(Q^2)}{2\pi}\mc C_r^{F(1)}(x)\right)\otimes\left[\sum_F \lm_B^F\left\{D^h_{q_f}(x,Q^2)+D^h_{\bar q_f}(x,Q^2)\right\}\right]\nn\\
&& + \frac{\al_s(Q^2)}{2\pi}\mc C_r^{G(1)}(x) \otimes \lm_B^G D^h_G(x,Q^2),
\eea
From the above equation it is clear that at NLO, the presence of gluon
term in the cross-sec equation contributes explicitly 
in the beginning itself at low $Q^2$ (for LO case it contributes
only through evolution of $D_q$'s). For $r=1,2$, the fragmentation functions
(FFs) $D^h_P(x,Q^2)$, dependent on $Q^2$, are written explicitly at NLO
as
\bea
f^h_{B,r}(x,Q^2) &=& \sum_F \lm_B^F\int_x^1 \frac{dz}{z} \left(\dl(1-z)+\frac{\al_s(Q^2)}{2\pi}\mc C_r^{F(1)}(z)\right) \left\{D^h_{q_f}\left(\frac{x}{z},Q^2\right)+D^h_{\bar q_f}\left(\frac{x}{z},Q^2\right)\right\}\nn\\
&& +  \lm_B^G \frac{\al_s(Q^2)}{2\pi}\int_x^1 \frac{dz}{z}\mc C_r^{G(1)}(z)D^h_G\left(\frac{x}{z},Q^2\right).
\eea
Hence,
\bea
\frac{d\s^{h}_{e^+e^-}(x;Q^2)}{dx} &=& \left(\frac{N_c4\pi\al^2}{3Q^2}\right)
\left[ \frac{3}{2}f^h_{B,1}(x,Q^2) - \frac{1}{2}f^h_{B,2}(x,Q^2)\right],\nn \\
\by{1}{\s_{tot}} \left(\frac{d\s^{h}_{e^+e^-}(x;Q^2)}{dx}\right)
&=& \frac{1}{\sum_F\lm_B^F \left(1 + \frac{\al_s}{\pi}\right)} \left[ f^h_{B}(x,Q^2)\right],
\eea
where,
\bea
f^h_{B}(x,Q^2) &=&  \int_x^1 \frac{dz}{z}\left[ \sum_F \lm_B^F\left(\dl(1-z)+\frac{\al_s(Q^2)}{2\pi} \mc C^{F(1)}(z) \right) \left\{D^h_{q_f} +D^h_{\bar q_f}\right\}\left(\frac xz\right)\right.\nn\\&&\left.+ \frac{\al_s(Q^2)}{2\pi}\lm_B^G\mc C^{G(1)}(z)D^h_G\left(\frac xz\right)\right],\\
\s_{tot} &=& N_c \sum_F{\lm_B^F}{\left(\frac{4\pi\al^2}{3Q^2}\right)}
\left( 1 + \frac{\al_s}{\pi}\right),  \hbox{ at  NLO.}
\label{eq:signlo}
\eea
Eq.~(\ref{eq:signlo}) contains the complete set of equations for
cross-section upto next-to-leading order level.
The expressions for co-efficient functions $\mc C^{F(1)}(z)$ and 
$\mc C^{G(1)}(z)$ of Eq.~(\ref{eq:signlo}) are taken 
from Appendix II of \cite{Fur}.

\paragraph*{Perturbative Functions}:
The evolution of the parton fragmentation function $D_i(x, Q^2)$
with $Q^2$ is given by DGLAP evolution equations \cite{Dglap}:

\bea
Q^2\frac{d}{dQ^2}D^h_P(x,Q^2) =\sum_{j}\mc P_{ji}(x,\al_s) \otimes D^h_{j}\left(x,Q^2\right)= \sum_{j} \int_x^1 \frac{dz}{z} D^h_{j}\left(\frac xz,Q^2\right) \mc P_{ji}(z,\al_s),
\eea
where $\mc P_{ji}$'s are the time-like splitting functions and can be expressed as power series in $\al_s$ \cite{Ams}:
\bea
\mc P_{ji}(x, \al_s) = \frac{\al_s}{2\pi}\mc P_{ji}^{(0)}(x) + \left(\frac{\al_s}{2\pi}\right)^2\mc P_{ji}^{(1)}(x)+\ldots .
\eea
Expressions for $\al_s$ and time-like splitting functions to NLO are
taken from Refs.~\cite{Fur},\cite{Petro} and \cite{Ellis} where they were
discussed in detail. The main difference between LO and NLO cross-sections
are perturbative expansion of functions like $\al(x, Q^2)$ and time like
splitting functions $P_{ji}$ which appears in the evolution equations.
The expression for first term in Eq.~(\ref{eq:signlo}) 
 gives the LO term \cite{Ams}.
\bea
\by{1}{\sigma_{tot}}\by{\d \sigma^h}{\d x} & = & \by{\sum_q c_q \,
D_q^h(x, Q^2)}{\sum_q c_q}~,
\label{eq:siglo}
\eea
where $D_q^h(x,Q^2)$ are analogously evolved to LO as well.

Eq.~(\ref{eq:siglo}) can be written in terms of singlet and non-singlet
combinations including the co-efficient functions \cite{Savee}
having three light quarks ($u, d$ and $s$) at the starting scale and
evolved to the $Z$-pole with the heavy quarks $c$ and $b$
contributing during the evolution. 

Vector mesons production of $\rh$ and $\om$s has been studied for LEP
data \cite{Rho},\cite{Omega}. However, in the case of $K^*$ and $\ph$
instead of LEP \cite{Kstar},\cite{Phi}, SLD pure ``uds" data (three flavors
alone) \cite{SLD} is used in order to avoid the contamination of heavy
flavor mesons decay in the data as they decay preferably to one of these
two strange mesons since $\vert V_{cs}\vert$ is large.

\section{Model for Vector Meson Fragmentation}

This section gives the outline of the broken $SU(3)$ model that we will
use in our analysis. Detailed explanations can
be seen in our earlier work \cite{Savee}.

Perturbative QCD evolves these fragmentation
functions through DGLAP evolution equations \cite{Dglap} to 
say, upto $Z$-pole, once they are defined by means of
parameterization at a starting energy scale, $Q_0^2 = 1.5$GeV$^2$.

The fragmentation functions can be parameterized by comparison with
the experimental data. A model is needed to determine
these fragmentation functions and we use a model with broken $SU(3)$
\cite{Savee}. In this paper, the fragmentation of the
vector mesons $\rh(\rh^+, \rh^-, \rh^0)$, $K^*(K^{*+}, K^{*-}, K^{*0},
\overline{K}^{*0})$, $\om$ and $\ph$ is studied. For each meson, six
quark and anti-quark fragmentation functions, $D_f^h(x,Q^2)$,
$f=u,d,s,\overline{u}, \overline{d}, overline{s}$, and a gluon
fragmentation function $D_g^h(x,Q^2$ associated with the meson
production, needs to be described at the starting scale.

Though $SU(3)$ flavor symmetry is not an exact symmetry, still it holds
good in describing the octet of vector mesons with a symmetry breaking
parameter. This model succeeded in explaining the octet mesons at
LO in QCD with appreciable results for both $e^+\,e^-$ and $p\,p$
scattering \cite{Savee}.

To explain the structure and purpose of the model let us begin with 
applying the $SU(3)$ symmetry for a general process:   
$$
q_i \rightarrow h^i_j + X_j~.
$$

Under $SU(3)$ this will be written as $3 \rightarrow 8 + X$ where $8$
represents the octet hadron and $X$ are the debris that comes out along
with hadron. This implies that a quark fragment into an octet hadron
such that the possibilities for $X$ are either triplet $(3)$, antisixplet
$(\overline{6})$ or fifteenplet $(15)$. Let $\al(x,Q^2)$, $\bt(x,Q^2)$
and $\g(x,Q^2)$ be the corresponding unknown SU(3) symmetric independent
fragmentation functions for each of these possibilities \cite{Indu},
that is, for X to be $3$  the probability of the quark to fragment into
an octet hadron is $\al$. In a similar way, the other two 
$X$ (= $\overline{6}$, $15$) values corresponds to the following 
fragmentation functions $\bt$ and $\g$.

Likewise, an anti-quark also produces an octet hadron with
$X$ being an anti-triplet ($\overline{3}$), sixplet (6) or
anti-fifteenplet ($\overline{15}$), for which $\overline{\al}(x,Q^2)$,
$\overline{\bt}(x,Q^2)$ and $\overline{\g}(x,Q^2)$ have to be
determined.

To study the meson octet, we have a total of 56 $(8 \times 7)$ unknown
fragmentation functions that has to be fitted with the data which is not
an easier case to deal. Thus introducing this $SU(3)$ symmetry reduces
the complexity of $56$ unknown fragmentation functions into just seven
fragmentation functions for all the mesons in the octet. This is further
reduced by charge conjugation symmetry to just four, $\alpha, \beta,
\gamma$ and the gluon, $D_g$.

\subsection{Representation of Fragmentation Functions}

All the octet mesons are represented by means of the above mentioned
three independent fragmentation functions given in Table~\ref{tab:frag}.
In addition to the symmetry property, {\it{isospin and charge conjugation
invariance}} of vector mesons $\rh(\rh^+, \rh^-, \rh^0)$, $K^*(K^{*+},
K^{*-}, K^{*0}, \overline{K}^{*0})$ reduces the three independent
unknown quark fragmentation functions further into functions named valence
($V$) and sea ($\g$). Assuming sea as flavor symmetric, the functions
can be written as follows:

\begin{equation} 
V(x, Q^2) = \al(x,Q^2) - {3 \over 4 }\g(x,Q^2)~,
\label{eq:val}
\end{equation}

\begin{equation}
S(x,Q^2) = 2 \g(x,Q^2)~.
\label{eq:sea}
\end{equation}
 
Eqs~.(\ref{eq:val}) and (\ref{eq:sea}) are the two well-defined 
expressions which
describes the quark fragmentation function for all the mesons in the
octet that are produced.   

\paragraph*{Breaking Parameter}:
Since $SU(3)$ is not a good description of octet, a $x$-independent
parameter $\lm$ is introduced in order to explain the strangeness
suppression for $K^*$ meson. As the sea is flavor symmetric, suppression
is common for all the light quarks either $u, d$ or $s$. Meanwhile,
the $u$ and $d$ light quarks need suppression factor $\lm$ in
order to pick up a massive strange quark $s$ from sea. But in the case
of strange quark it does not need such factor as it will easily pick up
the other two quarks from the sea.

\section{Initial Parameterization}

The inputs of valence $V(x, Q^2)$, sea $\g(x, Q^2)$ and gluon $D_g(x,
Q^2)$ fragmentation functions at a starting scale of $Q_0^2 = 1.5$GeV$^2$
can be parameterized by means of a standard polynomial,

\begin{equation}
F_i(x) = a_i  x^{b_i}(1-x)^{c_i}(1 + d_i x + e_i x^2)~,
\label{eq:func}
\end{equation}

where $a_i$, $b_i$, $c_i$, $d_i$ and $e_i$ are the values to be
determined.  The contribution of heavy quark flavors are zero at the
starting scale, but they are added in the appropriate thresholds during
evolution.

\section{Meson Fragmentation}
\subsection{Pure octet}
\paragraph*{Combination of functions for $\rh$ and $K^*$}:
The singlet combination of valence and sea fragmentation for one of
the $\rh$ and $K^*$'s are given as (for more details, see
Ref.~\cite{Savee}):
\bea
D_0^{{\rho}^+} =\ D_{u+{\overline{u}}+d+{\overline{d}}+
                    s+{\overline{s}}}^{{\rh}^+} \quad
               =\ 2V + 12\g~,
\label{eq:D0rho}
\eea
\begin{equation}
D_0^{K^{*+}}  =\  D_{u+{\overline{u}}+d+{\overline{d}}+
                    s+{\overline{s}}}^{K^{*+}} \quad
              =\  (1 + \lm)V + 12\lm\g~.
\label{eq:D0Kstar}
\end{equation}
The term $\lm$ in Eq~.(\ref{eq:D0Kstar}) refers to the strangeness
suppression parameter. We can also write the non-singlet combinations in
the same way. In addition to this, a suppression factor $f^{K^*}$ with
$D^{K^*}_g = f^{K^*}_g D^{\rho}_g$ for gluons is introduced.

\subsection{Mixture of octet and singlet}
The broken $SU(3)$ model is now extended to $\om$ and $\ph$ mesons
which are orthogonal combinations of the $SU(3)$ octet ($\om_8$) and
singlet states ($\om_1$). 
\bea \nn
\om & = & \sin\tta~\om_8 + \cos\tta~\om_1~, \\
\ph   & = & \cos\tta~\om_8 - \sin\tta~\om_1~,
\label{eq:statedef}
\eea
where $\omega_8 = (u\overline{u} + d\overline{d} -
2s\overline{s})/\sqrt{6}$, $\omega_1 = (u\overline{u} + d\overline{d}
+ s\overline{s})/\sqrt{3}$ are the corresponding orthogonal states and
$\tta$ is the vector mixing angle. 

\subsubsection{Singlet hadron ($\om_1$) fragmentation}
Consider the same process which is discussed earlier in the octet case. 
$$
q_i \rightarrow h + X_i~,
$$
in which a quark hadronises into a singlet meson so that $X$ can only
be a triplet ($3 \rightarrow 1 + X$). Therefore, we need to determine
only one unknown fragmentation function $\dl(x, Q^2)$ in the singlet
case. It is already known that the probability of a quark fragmenting 
into an hadron with $X$ being triplet is $\al(x, Q^2)$.
Hence we use the simple ansatz that the function $\dl$ is simply 
related to $\al(x, Q^2)$, the fragmentation function for 
members of octet meson. 
\begin{equation}
\frac{\dl}{3} = \frac{f_1 \,\al}{3} = \frac{f_1}{3}\left(V +
\frac{3}{4}\g\right)~,
\label{eq:dl}
\end{equation}
where the factor $1/3$ is normalisation term and the proportionality
constant parameter $f_1$ has to be determined in the analysis. Having 
both octet and singlet terms, the expressions for 
singlet and non-singlet combinations are presented further below.
 
\subsubsection{Combination of functions for $\om$ and $\ph$}
\paragraph{Octet part}: Let us begin with $\om_8$ as it
falls under pure $SU(3)$ octet. The fragmentation functions for light
quarks ($u$, $d$ and $s$) given in the Table~\ref{tab:frag} can be
written as follows:
\bea
D_u^8 &=& \frac{V}{6} + 2 f_{\rm sea} \g~; \\ \nonumber
D_s^8 &=& \frac{2}{3} \lambda V + 2 f_{\rm sea}\g~.
\label{eq:du8}
\eea
The new parameter $f_{\rm sea}$ is the unknown suppression factor for
the SU(3)-symmetric sea fragmentation functions; however the other terms
follow the same definitions.

\paragraph{Singlet part}:
Using our ansatz for the singlet hadron, we have
\begin{eqnarray}
D_u^1  = D_d^1 &=& \frac{f_1^u}{3} \left( V + 
\frac{3}{4} f_{\rm sea} \g \right)~, \\ \nonumber
D_s^1  &=& \frac{f_1^s}{3} \left( \lambda V + 
\frac{3}{4} f_{\rm sea} \g \right)~.
\label{eq:du1}
\end{eqnarray}
The suppression factors for the $u, d$-
and $s$-type singlet fragmentation functions are also to be determined 
from this analysis. In general, the expression for the fragmentation 
functions with vetor mixing angle $\tta$ are 

\begin{eqnarray} 
D_i^\phi &=& (c_i^\phi)^2\left(\cos^2\tta \by{D_i^8}{(c_i^8)^2} +
\sin^2\tta \by{D_i^1}{(c_i^1)^2}\right); \\ \nonumber
D_i^\omega &=& (c_i^\omega)^2\left(\sin^2\tta \by{D_i^8}{(c_i^8)^2} +
\cos^2\tta \by{D_i^1}{(c_i^1)^2}\right)~.
\label{eq:duop}
\end{eqnarray}

the term $i$ refers to the three light quarks ($u$, $d$ and $s$). The 
co-efficient function values are taken from \cite{Savee}. 
As usual, we again parameterize the gluon fragmentation functions 
as $D^{\omega,\phi}_g = f^{\omega,\phi}_g D^{\rho}_g$ and have to 
determine their values from comparison with data.

\section{Data Analysis and Results}

We started with the initial values for the parameters of fragmentation
functions and other scale independent functions that were used at the
leading order level (see Table~\ref{tab:inputs} and \ref{tab:unknown_par})
in \cite{Savee}. All the NLO terms are defined in the appropriate places
at the starting scale of $Q_0^2 = 1.5$ GeV$^2$ for only three light
quarks ($u, d, s$), whereas the heavy quarks are kept zero initially. The
broken $SU(3)$ model describes well in a simple manner about the three
light quarks at this low input scale. However when $Q^2$ evolves upto
the $Z$- pole the heavy quark contributions are added in the appropriate
thresholds such that they contribute during the evolution. The expression
for splitting functions, co-efficient functions, and running coupling
constant are defined in the NLO cross-section. The evolution code uses
an $x$-space evolution algorithm that is fast and efficient.

Analysis is done with the LEP data[\cite{Rho},\cite{Omega}] at $Z$-pole
$(Q^2)=(91.2)^2$ GeV$^2$ for $\rh$ and $\om$ mesons.  In the case of $K^*$
and $\ph$ instead of using LEP data [\cite{Kstar},\cite{Phi}], SLD pure
uds data (three flavors alone) \cite{SLD} are being used. The purpose of
using SLD data for strange mesons $K^*$ and $\ph$ is in order to avoid the
contamination of heavy mesons, like $B$ and $D$, decaying into one of the
strange mesons which will mix up with the direct fragmentation of a quark
into $K^*$ and $\ph$. Whereas in the case of non-strange mesons $\rh$
and $\om$ the $K$ meson decays into the least massive meson like $\pi$
rather than $\rho$ or $\omega$; hence contamination from heavy quarks
in these channels is negligible.

\subsection{Analysis of data}

Now, with the usual definitions for valence $V(x, Q^2)$, sea $\g(x,
Q^)$ and gluon g$(x, Q^2)$ fragmentation functions, and with the help of
Eqs.~(\ref{eq:D0rho}) and (\ref{eq:D0Kstar}), we fitted the data to $\rh$
and $K^*$ for $x$ values ranging from 0.01--1.  In general, the large $x$
behaviour is explained by valence $V(x, Q^2)$ and small $x$ behaviour by
sea $\g(x, Q^2)$.  The intermediate $x$ part is governed by both sea and
gluons which was not very well determined at LO in $e^+\,e^-$ scattering
\cite{Savee}.  Since there were no gluon terms in LO cross-section the
idea about gluon behaviour is completely ill-determined at that point,
whereas in NLO it is expected to get reasonable values of gluon related
fragmentation functions as well as constant parameters in the polynomial.

In the analysis, the statistical and systematical error bars were added
in quadrature. Since the data at specific $x$ values quite differ from
that averaged over $x$ bin values, the cross-section is calculated by
averaging over $x$ bins as per the data.  We have used $x_p$ values
throughout the analysis where $p$ refers to the momentum scale (that is,
$x \equiv 2E/\sqrt{s}$ while $x_p \equiv 2p\sqrt{s}$).

The valence and sea fragmentation functions parameterization are
first determined with only $\rh$ and $K^*$ because, in principle,
the valence contribution can be clearly predicted by pure non-strange
$\rh$ meson and the sea part (including suppression factor $\lambda$
by $\rho$ and the strange meson $K^*$. Hence, while determining these
parameters the $x$-independent parameter $\lm$ which explains the
strangeness suppression was also included in the evolution. After this
the parameterization for gluons was also included and the data for all
nonet mesons was simultaneously fitted, keeping all parameters including
those for valence and sea, free, and refitting all parameters together.

\begin{figure}[tph]
\begin{center}
\includegraphics[angle=-90, width=0.49\textwidth]{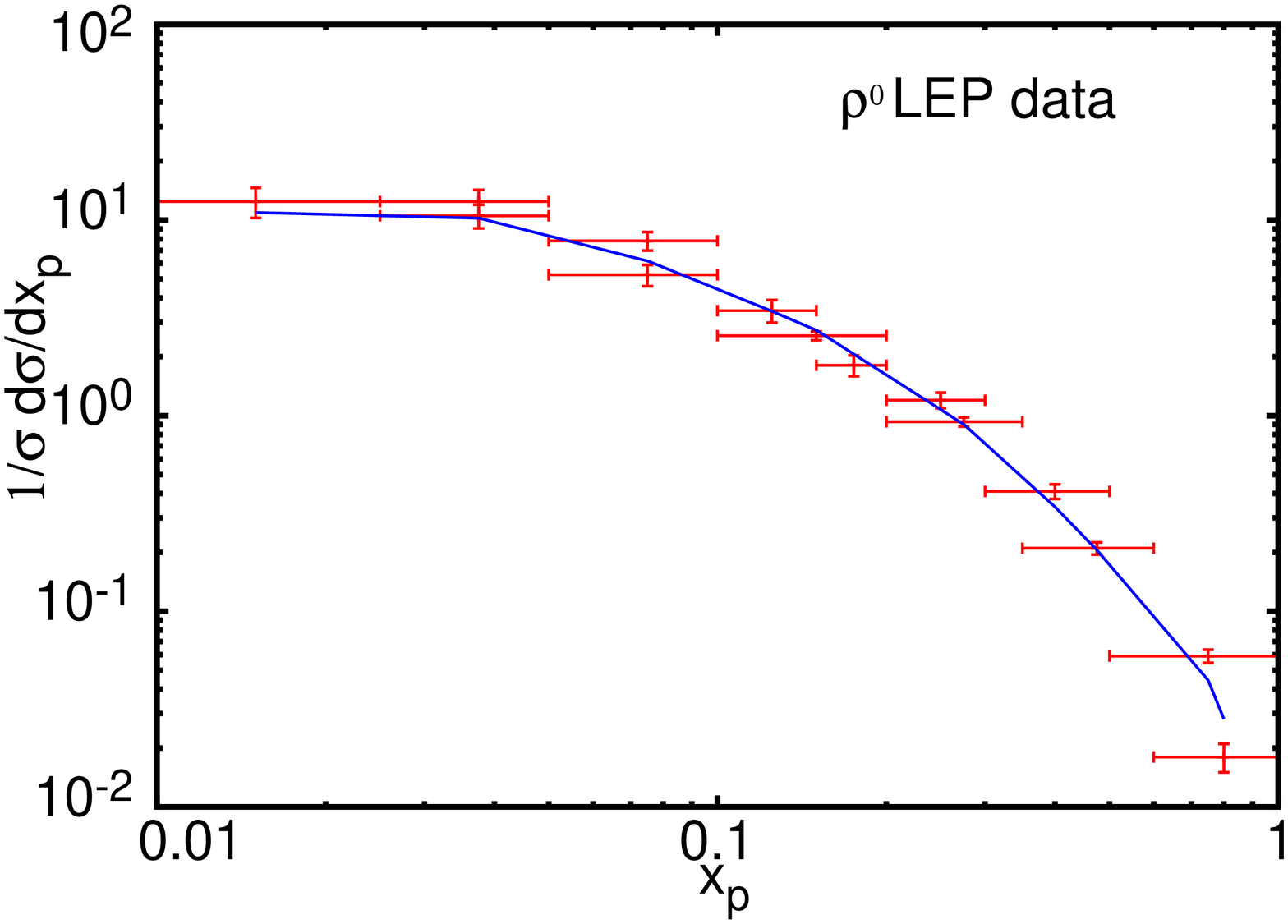}
\includegraphics[angle=-90, width=0.49\textwidth]{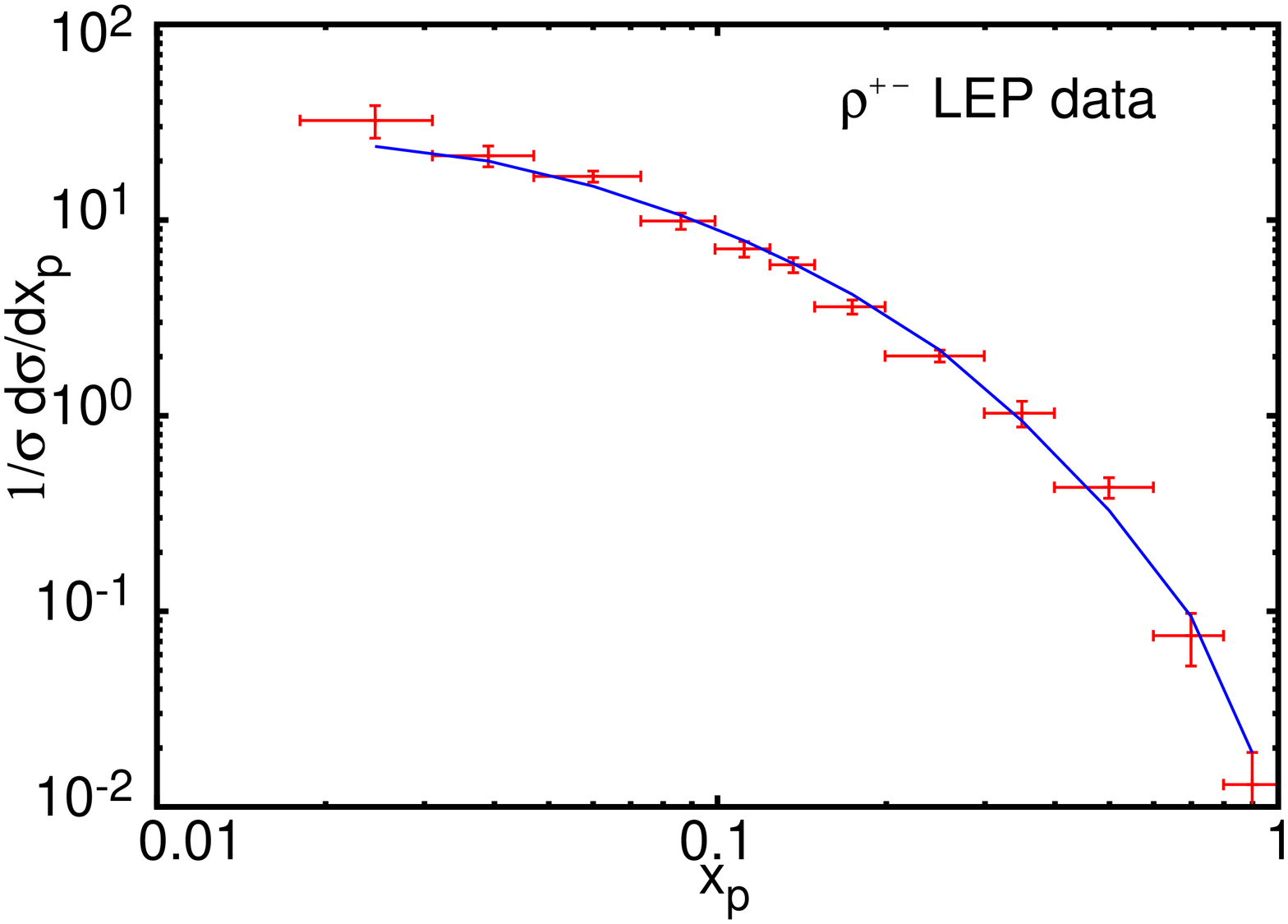}
\caption{Fit for rho meson in terms of fragmentation functions with
(L) LEP data on the $Z$-pole for two data sets (ALEPH-thick line and
DELPHI-thin line) of $\rh^{0}$ and (R) for $\rh^{+-}$ 
\cite{Rho}.  The data are shown with statistical and systematic errors
added in quadrature.}
\label{fig:rho}
\end{center}
\end{figure}

\begin{figure}[htp]
\begin{center}
\includegraphics[angle=-90,width=0.6\textwidth]{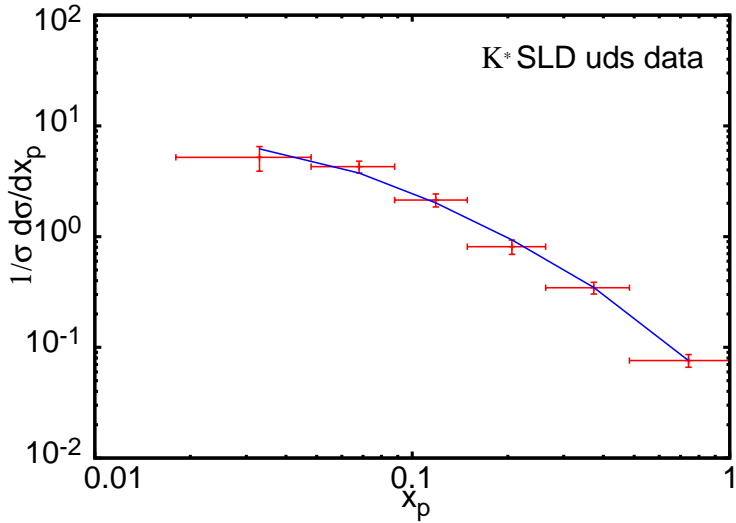}
\caption{Fit for $K^*$ meson with the best fit value of suppression factor 
$\lm = 0.07$. Data are taken from Ref.~\cite{SLD} at the $Z$-pole, 
from light quarks only and the smooth line refers the cross section at NLO.}
\label{fig:kstar}
\end{center}
\end{figure}

Fig. \ref{fig:rho} shows the result of the fit to both $\rho^{+,-}$
as well as $\rho^0$. Note that isospin symmetry implies $D_q^{\rho
+} + D_q^{\rho -} = 2 D_q^{\rho 0}$. Fig.~\ref{fig:kstar} shows
the distribution for SLD ``pure uds" data \cite{SLD} of $K^{*0} +
\overline{K}^{*0}$.

The main observation here is that after the minimization process we get
the value of $\lm$, suppression parameter, as $\lm = 0.07 \pm 0.01$. This
is roughly the same as the value of $\lm = 0.063\pm 0.01$ at LO level
obtained in Ref.~\cite{Savee}. So $\lm$ lies in the range $0.05 \le \lm
\ge 0.07$ with $1\s$ error bar (see Table~\ref{tab:unknown_par}). This
result again is close to the value $\lm = 0.08$ for pseudoscalar
mesons \cite{Misra} which implies that the $x$ independent suppression
parameter is completely spin-independent quantity. The sea suppression
factor $f_g^{K*}$ for strange meson  $K^*$ came out to be $f_g^{K*} =
1$, indicating no suppression. The $\chi^2$ values for the fit to both
$\rho$ and $K^*$ data is given in Table~(\ref{tab:chi2}).

In the case of $\omega$ and $\phi$ mesons, there are a few additional
unknown parameters like $f_1^u, f_1^s$ and $f_{sea}$ for $\om$ and $\ph$
which represents the singlet constants and sea suppression and mixing
angle $\tta$ are to be determined. Since $\om$ is highly dominated by $u$
and $d$ quarks it will behave as a purely non-strange meson, while $\phi$
is almost purely dominated by its $\s\overline{s}$ component.

\begin{figure}
\begin{center} 
\includegraphics[angle=-90, width=0.49\textwidth]{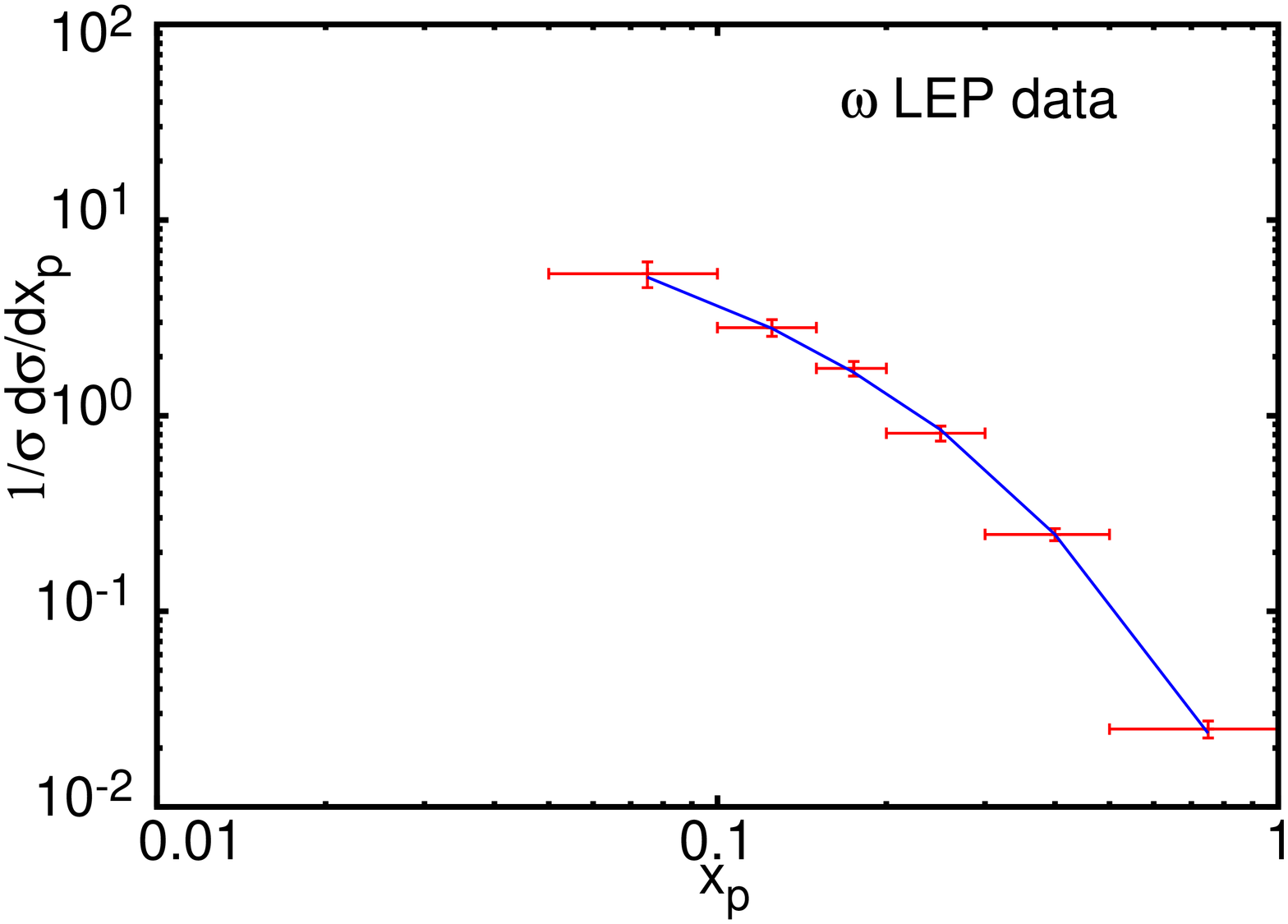}
\includegraphics[angle=-90, width=0.49\textwidth]{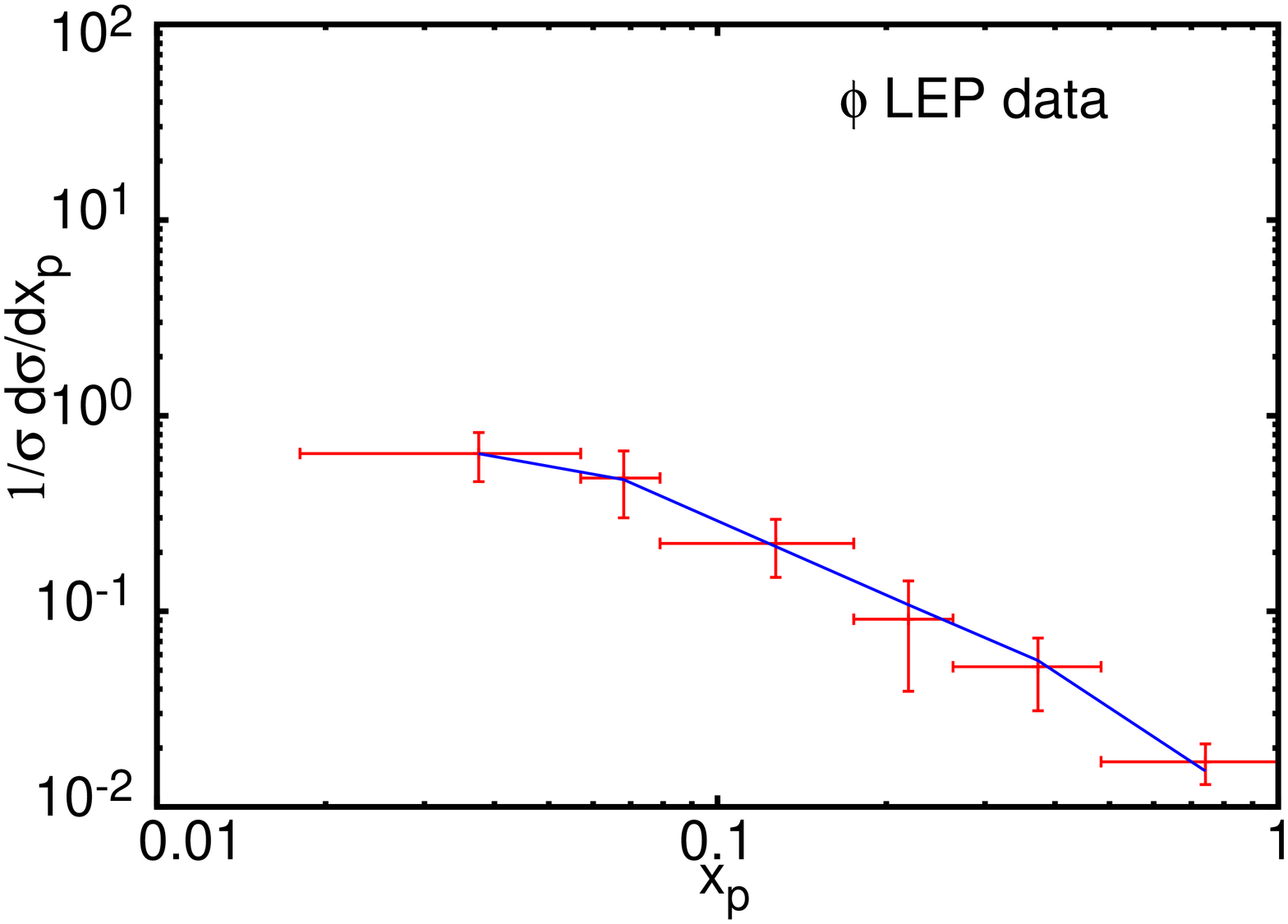}
\caption{Fits to omega (L) and (R) phi meson.
The data correspond to LEP data for $\omega$ \cite{Omega} and the SLD
data from light quarks alone \cite{SLD} for $\phi$ at NLO level.}
\label{fig:omegaphi}
\end{center}
\end{figure}

The comparison of the best fit to the $\omega$ and $\phi$ data are
shown in the Fig.~\ref{fig:omegaphi}. The $\ph$ meson, an interesting
candidate for QGP studies, in the entire meson nonet, is also fitted to
only light quark pure uds SLD data \cite{Phi} as in the $K^*$ case.

The best-fit value of $\tta$, the vector mixing angle, is $\tta =
39$--42$^\circ$ at $1\sigma$. This value is reasonably close to the
value of $\tta =42\pm 2^\circ$ obtained in the earlier LO analysis
\cite{Savee}. This is also reasonably close to the value $\tta =
36^{\circ}$--$38.7^\circ$ \cite{PDG}.

As discussed earlier in \cite{Savee}, the values of the constants
$f_1^s$ of $\om$ meson and $f_1^u$ of $\ph$ are kept zero. The other
two constants are $f_1^u = 0.05$, $f_{sea}^{\om} =0.99$ which are
still sensiblly close to the LO results and the constant factor for
gluon suppression value $f_g^{\om}=1$ with large errors.

The unknown constants $f_{sea}^{\phi}$, $f_1^s$ and $f_g^{\ph}$ of
$\ph$ meson are also determined by fitting with the data.  According to
Ref.~\cite{Yao} the  value of $\tta$ close to $35^\circ$ value saturates
the physical $\ph$ state as a pure $s\overline{s}$ state. Thus we fixed
the sea suppression factor for $\ph$ as $f_{sea}^{\phi} = \lm^2$ as it
contains dominantly strangeness $(s\overline{s})$ state, while $f_1^s=5.63$,
again with larger errors.

The gluon suppression parameter for $\phi$ is tightly constrained as
$f_g^\phi=0.4\pm 0.04$ in contrast to hardly any suppression required
in $\omega$.  Hence it is clearly understood that $\ph$ is a dominantly
strange meson with both sea and gluon contributions suppressed.

The best-fit values for all the parameters and their $1\sigma$ errors
are given in Table~(\ref{tab:unknown_par}. It can be seen that all the
polynomial constants in the three fragmentation functions are consistent
with zero. The $\chi^2$ values of the individual fits to the different
meson data is given in Table \ref{tab:chi2}.

The global $\chi^2$ for entire meson nonet was $17.1$ with $44$ data
points and $21$ degrees of freedom. This reflects $70$ percent confidence
level for the goodness-of-fit which looks reasonable.

Hence with the broken $SU(3)$ model we are able to explain the entire 
meson octet with the introduction of very few fragmentation fuctions 
basically and some additional $x$- and $Q^2$-independent parameters
for extension of the octet and singlet mixture at the next-to-leading
order level. This proves the efficiency of the model to explain the
sparse data with minimum number of fragmentation functions.

\section{Discussion and Conclusion}  

Fragmentation functions for the vector meson nonet ($\rh(\rh^+, \rh^-,
\rh^0)$, $K^*(K^{*+}, K^{*-}, K^{*0}, \overline{K}^{*0})$, $\om$ and
$\ph$) are studied for the first time at the next-to-leading order
level for $e^+ \, e^-$ scattering. To achieve this we used a simple
model named {\it{broken}} $SU(3)$, since $SU(3)$ is a fair description
of octet mesons.

The model uses $SU(3)$ symmetry along with charge conjugation and isospin
symmetry to describe the three input fragmentation functions $\al(x,
Q^2)$, $\bt(x,Q^2)$ and $\g(x,Q^2)$ (see Table~(\ref{tab:inputs})) at
an input scale of $Q_0^2 = 1.5$ (GeV)$^2$ for three light quarks $(u,
d, s)$ where the heavy flavors like charm and bottom quarks were zero
at this scale.  These, along with the gluon fragmentation functions,
were evolved to the $Q^2$ of the data.  The heavy quark contributions
were added in appropriate thresholds during the evolution.

With the standard input parameterization, these fragmentation
functions were evolved through DGLAP evolution equations at NLO and the
cross-sections were calculated for $\rh$, $K^*$ $\omega$ and $\phi$
mesons. The symmetry was broken when we introduced a $x$ independent
parameter $\lm$ for strange mesons to explain the strangeness suppression
for non-strange mesons to pick up a strange quark from sea.

The results obtained were fitted with the momentum scale of $\sqrt{s}=
91.2$ GeV for LEP ($\rh$ and $\omega$) and SLD data ($K^*$ and $\phi$)
at the NLO level and the fit looks reasonable at NLO. The best fit values
of the parameters for quark and gluon fragmentation functions ($V, \g,
g$) with the error bars were given in Table~(\ref{tab:inputs}). The
suppression parameter, $\lm = 0.07$ value obtained in NLO is almost
equal to the LO result and still close to pseudoscalar meson $\lm=0.08$
value. This clearly shows that $\lm$ is a spin independent factor.

The values of the gluon fragmentation functions were well determined
in this study. The behaviour both at small $x$ ($x^b$ dependence) and
at large $x$ ($(1-x)^c$ dependence) in gluon fragmentation function
lies between Valence and Sea fragmentation functions. That is at
large $x$ the $\g(x, Q^2)$ falls first, then the gluon $g(x, Q^2)$,
and finally $V(x, Q^2)$ towards the largest $x$ values. At LO, the
gluon fragmentation appears only in the DGLAP evolution equation. The
NLO cross sections, in contrast to the LO case, are directly dependent
on the gluon fragmentation function. This is reflected in the tight
constraints on the gluon fragmentation function parameters at NLO as
seen in Table \ref{tab:inputs}. In particular, the fits are very
sensitive to the small-$x$ behaviour of the gluon, which transforms from
a very converging, vanishing function at low $Q^2$ to a highly diverging
one at larger $Q^2$ due to the poles at $x=0$ in the relevant splitting
functions. Specifically, it can be seen that the exponent $b$ in $x^b$
is extremely well-determined for the gluon fragmentation function.

In addition, it is seen that the gluon fragmentation functions are
severely suppressed in $\phi$ but not in $\omega$, which are mixtures
of SU(3) octet and singlet mesons. This is also reflected in the
value of the singlet-octet mixing parameter $\tta$, close to $\tta
\sim 41^\circ$. The values of the constants simply implies that $\om$
is dominantly a non-strange meson and $\ph$ is dominantly a strange meson.

All the parameter values of quark and gluon fragmentation functions
with the additional constant parameters with error bars were given
in Tables~(\ref{tab:inputs}) and (\ref{tab:unknown_par}). Finally the
$\chi^2$ values for each meson were tabulated Table~(\ref{tab:chi2}).

In summary, we got reasonable fits for all the mesons at the NLO level for
$e^+\,e^-$ scattering which indeed implies the consistency and efficiency of
this model that explains the fragmentation functions of vector mesons
both at the leading and the next to leading order in QCD. This work will
be extended further to $p\,p$ collision in future to understand further
the $\ph$ meson in particular which forms the baseline to understand
the QGP.

\paragraph{Acknowledgements}: HS thanks A S Vytheeswaran for constant
encouragement and support and M V N Murthy for discussions.

\newpage
\begin{table}
\centering
\begin{tabular}{|ccl|ccl|} \hline
fragmenting & \multicolumn{2}{c|}{${}_{\displaystyle K^{*+}}$} & fragmenting &
\multicolumn{2}{c|}{${}_{\displaystyle K^{*0}}$} \\
quark & & & quark & & \\  \hline
$u$ & : &  ${\al}+{\bt}+{\frac{3}{4}}{\g}$ & 
$u$ & : &  $2{\bt}+{\g}$ \\
$d$ & : &  $2{\bt}+{\g}$ & 
$d$ & : &  ${\al}+{\bt}+{\frac{3}{4}}{\g}$ \\
$s$ & : &  $2 {\g}$ & 
$s$ & : &  $2 {\g}$ \\ \hline
fragmenting & \multicolumn{2}{c|}{${}_{\displaystyle \omega/\phi}$} & 
fragmenting &
\multicolumn{2}{c|}{${}_{\displaystyle \rho^0}$} \\
quark & & & quark & & \\  \hline
$u$ & : &  
$\frac{1}{6}{\al}+\frac{9}{6}{\bt}+\frac{9}{8}{\g}$ &
$u$ & : &  
$\frac{1}{2}{\al}+\frac{1}{2}{\bt}+\frac{11}{8}{\g}$ \\
$d$ & : &  
$\frac{1}{6}{\al}+\frac{9}{6}{\bt}+\frac{9}{8}{\g}$ &
$d$ & : &  
$\frac{1}{2}{\al}+\frac{1}{2}{\bt}+\frac{11}{8}{\g}$ \\
$s$ & : &  
$\frac{4}{6}{\al}+\frac{9}{6}{\g}$ &
$s$ & : &  $2\bt+\g$ \\ \hline
fragmenting & \multicolumn{2}{c|}{${{}_{\displaystyle \rho^+}}$} &
fragmenting &  \multicolumn{2}{c|}{${{}_{\displaystyle \rho^-}}$} \\
quark & & & quark & & \\  \hline
$u$ &  : & ${\al}+{\bt}+{\frac{3}{4}}{\g}$ & 
$u$ &  : & $2 {\g}$ \\ 
$d$ &  : & $2 {\g}$ & 
$d$ &  : & ${\al}+{\bt}+{\frac{3}{4}}{\g}$ \\ 
$s$ &  : & $2{\bt}+{\g}$ & 
$s$ &  : & $2{\bt}+{\g}$ \\ \hline
fragmenting & \multicolumn{2}{c|}{${{}_{\displaystyle \overline{K^{*0}}}}$} &
fragmenting & \multicolumn{2}{c|}{${{}_{\displaystyle K^{*-}}}$} \\
quark & & & quark & & \\  \hline
$u$ & : & $2{\bt}+{\g}$ & 
$u$ & : & $2 {\g}$ \\ 
$d$ & : & $2 {\g}$ & 
$d$ & : & $2{\bt}+{\g}$ \\ 
$s$ & : & ${\al}+{\bt}+{\frac{3}{4}}{\g}$ & 
$s$ & : & ${\al}+{\bt}+{\frac{3}{4}}{\g}$ \\
\hline
\end{tabular}
\caption{Quark fragmentation functions into members of 
 meson octet in terms of the SU(3) functions, $\al$,
$\bt$ and $\g$. }
\label{tab:frag}
\end{table}
\vskip 2truecm
\begin{table}
\centering
\begin{tabular}{|c|c|c|c|c|} \hline
 &      & Central Value   & {Error Bars} \\ \hline
$V$     &$a$ & 0.72   & 0.09    \\ \hline
        &$b$ & 0.52   & 0.15    \\ \hline
        &$c$ & 1.24   & 0.10    \\ \hline
        &$d$ & 0.27   & 0.28    \\ \hline
        &$e$ & -0.16  & 0.14    \\ \hline
$\g$    &$a$ & 0.99   & 0.01    \\ \hline
        &$b$ & -0.48  & 0.02    \\ \hline
        &$c$ & 5.48   & 0.14    \\ \hline
        &$d$ & -0.09  & 0.18    \\ \hline
        &$e$ & 1.25   & 0.63    \\ \hline
$D_g$   &$a$ & 3.89   & 0.41    \\ \hline
        &$b$ & 0.745  & 0.001   \\ \hline
        &$c$ & 3.14   & 0.19    \\ \hline
        &$d$ & -0.13  & 0.23    \\ \hline
        &$e$ & -0.21  & 0.44    \\ \hline
\end{tabular}
\caption{Best fit values of the parameters for the input 
fragmentation functions at the initial scale of $Q^2 = 1.5 \ GeV^2$.}
\label{tab:inputs}
\end{table}
\vskip 2truecm
\begin{table}
\centering
\begin{tabular}{|c|c|c|c|c|} \hline
 &      Central Value     & {Error Bars} \\ \hline
$\lm$            & 0.07   & 0.01   \\ \hline
$\tta $          & 40.49  & 1.31   \\ \hline
$f_{sea}^{\om}$  & 0.99   & 0.08   \\ \hline
$f_1^u(\om)$     & 0.05   & 0.36   \\ \hline
$f_1^s(\ph)$     & 5.63   & 2.16   \\ \hline
$f_g^{K^*}$      & 1.00   & 0.09   \\ \hline
$f_g^{\om}$      & 1.00   & 0.38   \\ \hline
$f_g^{\ph}$      & 0.40   & 0.04   \\ \hline
\end{tabular}
\caption{Best fit values of the parameters for the input 
fragmentation functions at the initial scale of $Q^2 = 1.5 \ GeV^2$.}

\label{tab:unknown_par}
\end{table}
\vskip 2truecm
\begin{table}
\centering
\begin{tabular}{|c|c|c|c|c|} \hline
Data Set              & No. of data points  & $\chi^2$ \\ \hline
$\rh^0$ (ALEPH)       &    8                &     5.6   \\ \hline
$\rho^0$ (DELPHI '95) &    6                &     2.4    \\ \hline
$\rh^{+-}$ (OPAL)     &   12                &     6.1     \\ \hline
$K^{*0}$ (SLD)        &    6                &     2.9      \\ \hline
$\omega$ (ALEPH)      &    6                &     0.3       \\ \hline
$\phi$ (SLD)          &    6                &     0.2        \\ \hline
\end{tabular}
\caption{$\chi^2$ for fits to inclusive vector meson production
data on the $Z$-pole  from LEP and SLD experiments with a total of
$44$ data points and $23$ free parameters, we have a total 
$\chi^2 =17.1 $ for $44-23=21$ degrees of freedom.}
\label{tab:chi2}
\end{table}

\end{document}